\documentclass[fleqn,twoside]{article}
\usepackage{espcrc2}
\usepackage{graphicx}

\title{Charmonium properties at finite temperature on quenched
anisotropic lattices.}

\author{
Takashi Umeda%
    \address{Yukawa Institute for Theoretical Physics, 
             Kyoto University, Kyoto, 606-8502, Japan.}%
    \thanks{Talk presented by T. Umeda}
and
Hideo Matsufuru%
    \address{
             High Energy Accelerator Research Organization (KEK),
             Tsukuba, 305-0801, Japan}
}

\begin{document}

\begin{abstract}
We study charmonium properties
below and above $T_c$ up to 1.8$T_c$, on quenched anisotropic lattices.
Information of the spectral functions is extracted
using the maximum entropy method and the constrained curve fitting.
We also calculate the color singlet and averaged free energies
and evaluate the charmonium spectrum with the potential model analysis.
The relation between the lattice result of the spectral function
analysis and the potential model is discussed.
\vspace{1pc}
\end{abstract}

\maketitle

\section{Introduction}

To investigate the properties of quark gluon plasma (QGP)
in heavy ion collision experiments, theoretical prospects
are significant, since such processes include complicated
interactions among large number of particles.
Changes of charmonium states have been regarded as one of the most
important probes of plasma formation \cite{NA50},
because the potential model calculations predict the mass shift
of charmonium near $T_c$ \cite{Has86},
and $J/\psi$ suppression above $T_c$ \cite{Mat86}.
However, lattice QCD simulations have indicated that the
thermal properties of hadronic correlators are much more
involved than weakly interacting almost free quarks
\cite{TARO01}.
Recent studies of spectral functions of charmonium suggest that
a hadronic excitation of $c$-$\bar{c}$ system may
survives above $T_c$ \cite{Ume03,Pet03,Asa03}.
This seems to conflict with the predictions of
potential model approaches and a naive picture of QGP.

In this study we discuss the above disagreement between the
spectral function analysis and the potential model.
On one hand, we extract the information of the spectral function
from the temporal charmonium correlator.
On the other hand, we also extract the static quark potentials
from the color singlet Polyakov loop correlation and Wilson
loop.
The result of potential model calculation using the latter
is compared with the former.

These calculation are performed on quenched anisotropic lattices.
This enables simulations at temperatures from $T\sim 0$ ($Nt=160$)
to 1.75$T_c$ ($Nt=16$) keeping
the lattice cutoff unique, to avoid uncertainties caused
by the cutoff dependence in, for example, self-energy contribution
to the potentials.
The anisotropic lattice is also useful to keep the number of
data points of the temporal correlators sufficiently large,
which is important for detailed analysis of the spectral
functions.
We extend the simulation reported in Ref.~\cite{Ume03}
which was performed at the renormalized anisotropy 
$a_\sigma/a_\tau=4$,
the spatial lattice cutoff is $a_\sigma^{-1}=2.0$GeV, and
with the plaquette gauge and $O(a)$ improved Wilson quark actions.
The critical temperature almost corresponds to $N_t=28$.

\section{Spectral functions}

To extract the spectral function from the temporal correlators,
we adopt the maximum entropy method (MEM) and
the constrained curve fitting (CCF) \cite{Lep02}.
The result of the former provides the prior knowledge required
for the latter.
Combining the two methods, the results may be more
reliable and quantitative than with one of them.
In our previous study \cite{Ume03} we found that MEM fails to
extract the spectral function from the correlator of local
operators at $T>0$.
Therefore we adopt spatially extended operators to 
enhance the low frequency mode of the spectral function.
However the smeared operators may lead to an artificial peak,
and thus careful analysis is necessary to distinguish the
physical results from the artifact ones.

Although our MEM result shows that 
the spectral functions have peak structure in PS and V channels
(corresponding to $\eta_c$ and $J/\psi$) at all temperature, 
the difference of result for different smearing functions,
we call ``smeared'' and ``half-smeared'', exists at higher temperature,
especially above $1.4T_c$.
This means that the peak structure of the spectral function at high
temperature might be artificial.
Furthermore we find large default model function dependence of the
results, in which the position of the peak is stable but the peak width
has large dependence.
Therefore it is difficult to study the spectral function 
quantitatively using only MEM.

\begin{figure}[tb]
\includegraphics[width=70mm]{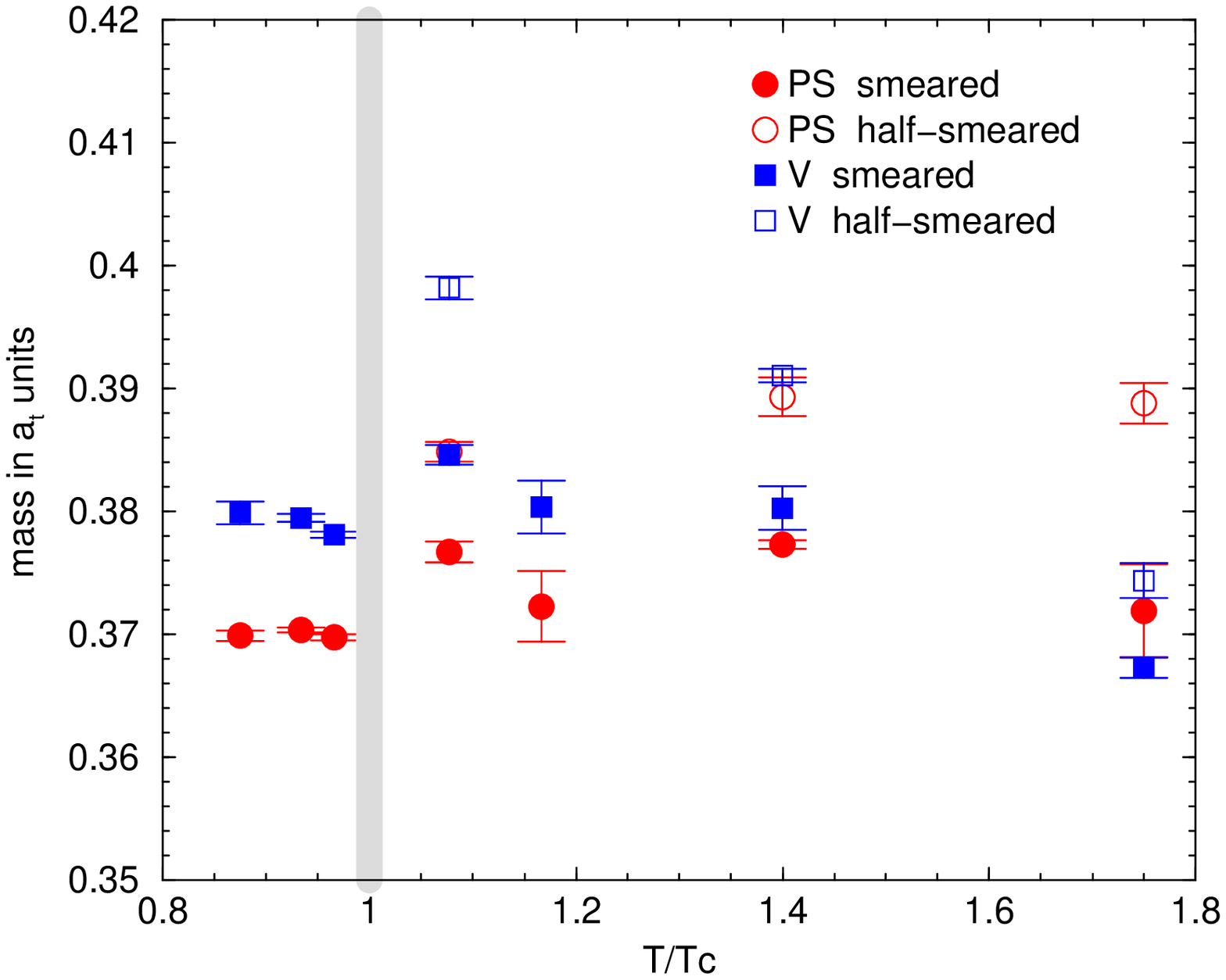}
\includegraphics[width=70mm]{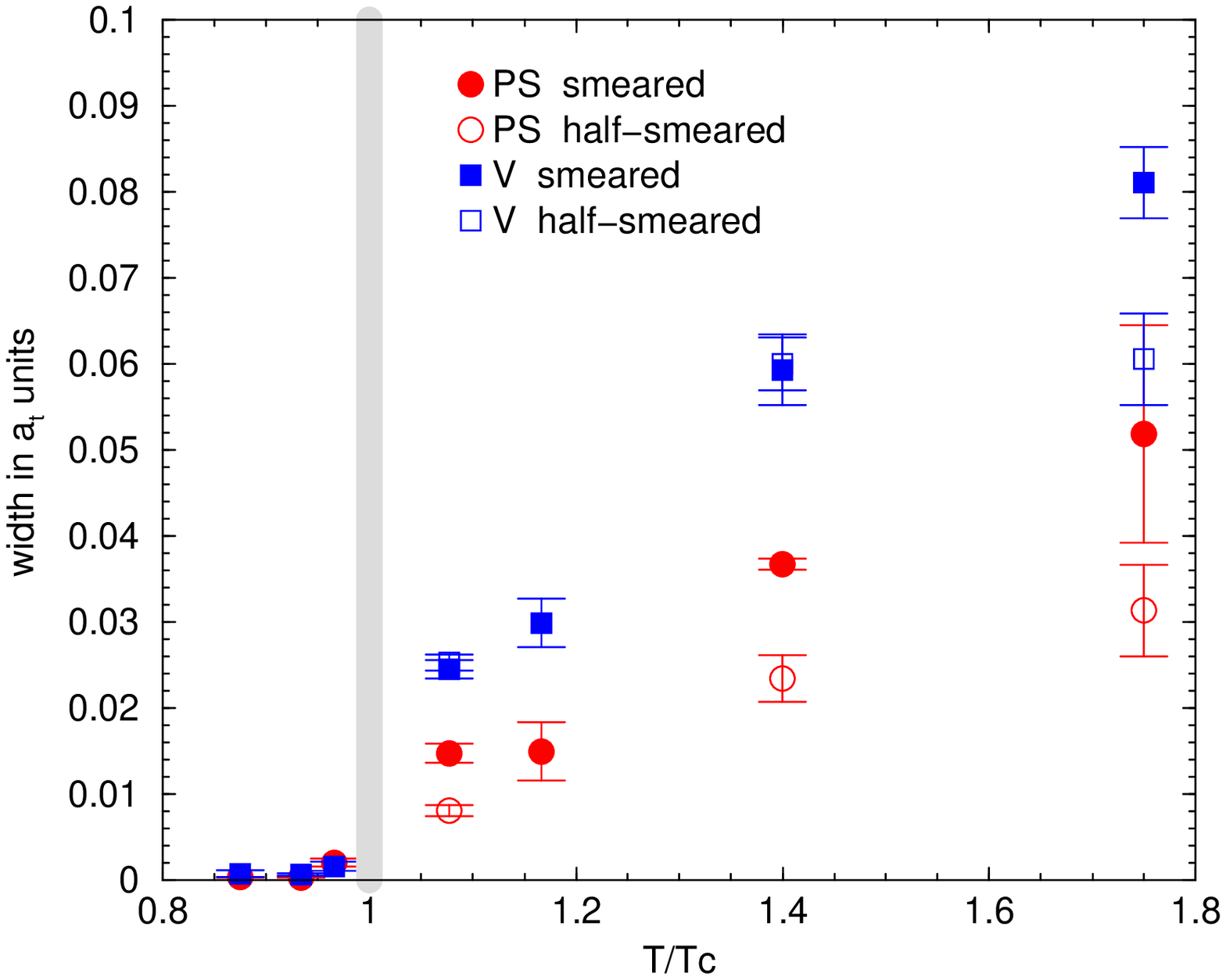}
\vspace{-11mm}
\caption{Temperature dependence of lowest peak mass, $m_i$, 
and width, $\gamma_1$, from the result of CCF analysis.}
\label{fig1}
\vspace{-8mm}
\end{figure}

CCF analysis is performed based on the result of MEM,
assuming the form of fitting function as the Breit-Wigner type,
\begin{eqnarray}
A(\omega)=\sum_{i=1}^{N_{term}}\frac{\omega^2m_i\gamma_i R_i}
{(\omega^2-m_i^2)^2+m_i^2\gamma_i^2},
\end{eqnarray} 
where $R_i$, $m_i$ and $\gamma_i$ are overlap, mass and width of the
$i$-th peak respectively, and $N_{term}=2$ is used in this analysis.
The result of lowest peak position (mass) and width are shown in 
Fig.\ref{fig1}.
Unfortunately, our results of the CCF is unstable against the prior
knowledge as inputs.
Therefore the systematic uncertainties will 
be larger than the quoted statistical error in Fig.\ref{fig1}.

In spite of large systematic uncertainty, we can find the
following tendency.
Below $T_c$, the spectral function has a peak with almost the same
mass at $T=0$ and vanishing width.
Above $T_c$, the peak stays at similar position as at $T<0$
while its width tends to grow as temperature increases.
As pointed out in the MEM analysis, however, the difference
between the results of smeared and half-smeared correlators
shows that a disappearance of the peak structure at higher $T$
cannot be excluded, especially in the V channel.

\section{Static quark free energies}

In this section we compare the results of spectral function with the
potential model calculation.
Here we define the following potentials.
\begin{eqnarray}
V_{ave}(r) &=& -\ln{\langle \mbox{Tr}L(\vec{r})
\mbox{Tr}L^\dagger(\vec{0})\rangle}/(N_ta_\tau) \\ 
V_{sing}(r) &=& -\ln{\langle \mbox{Tr}L(\vec{r}) 
L^\dagger(\vec{0})\rangle}/(N_ta_\tau) \\ 
V_{Wilson}(r) &\propto& -\ln{W(\vec{r},t)}/t
\end{eqnarray}
where $L(\vec{r})$, $W(\vec{r},t)$ are Polyakov and Wilson loops,
respectively.  
$V_{ave}$ is adopted in the former potential model calculations.
$V_{sing}$ can not define in gauge independent form, then
we calculate it in the Coulomb gauge, in which the $V_{sing}$ 
will be equivalent with a gauge invariant definition \cite{Phi02}.
We can also define the other color singlet potential using the
Wilson loops, $V_{wilson}$, which is calculated with the same way
as at zero temperature.
$V_{singlet}$ is thermal average of the energy spectrum of the color
singlet $q-\bar{q}$ system, while $V_{wilson loop}$ is the lowest one of
the spectrum.

\begin{figure}[tb]
\includegraphics[width=70mm]{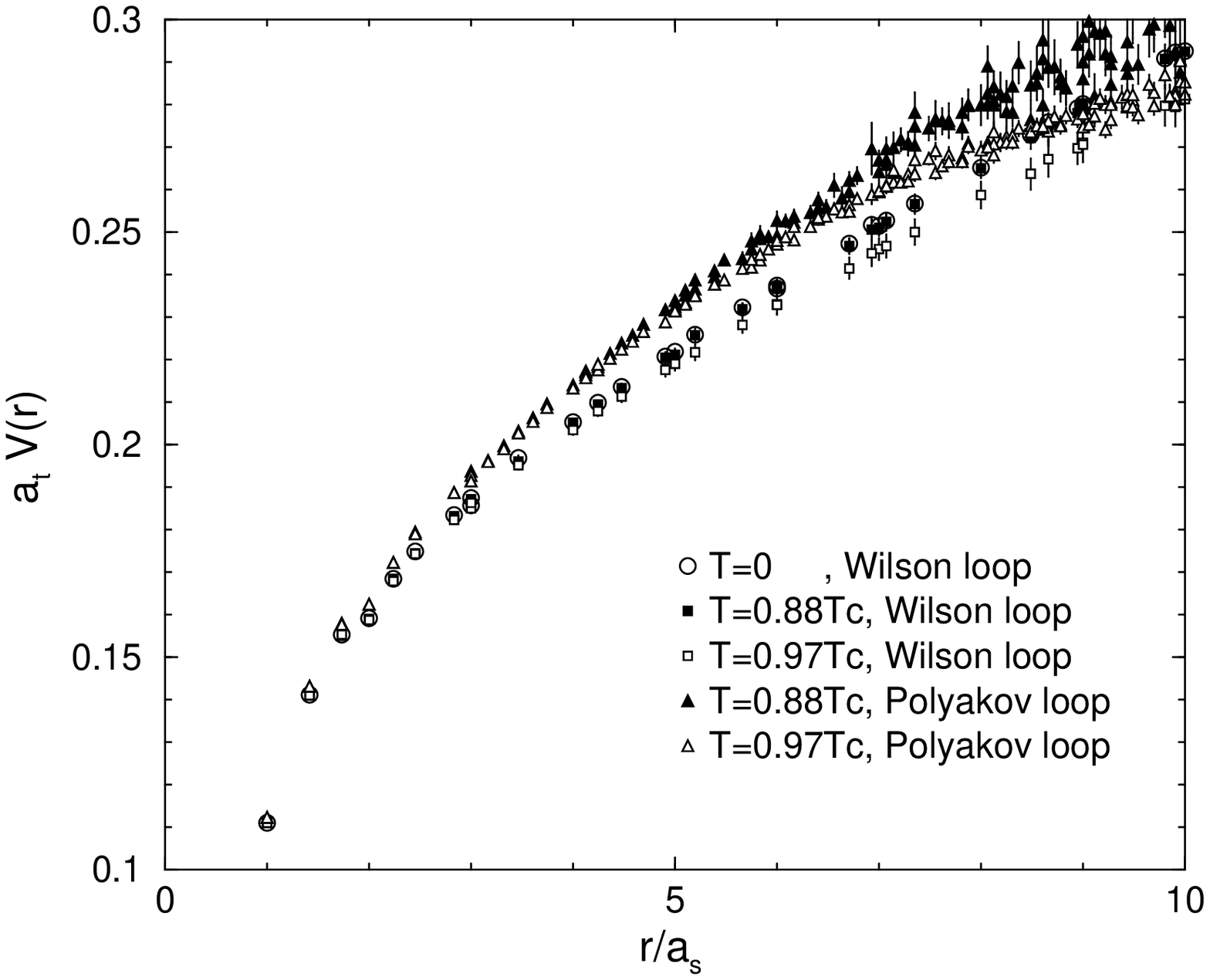}
\includegraphics[width=70mm]{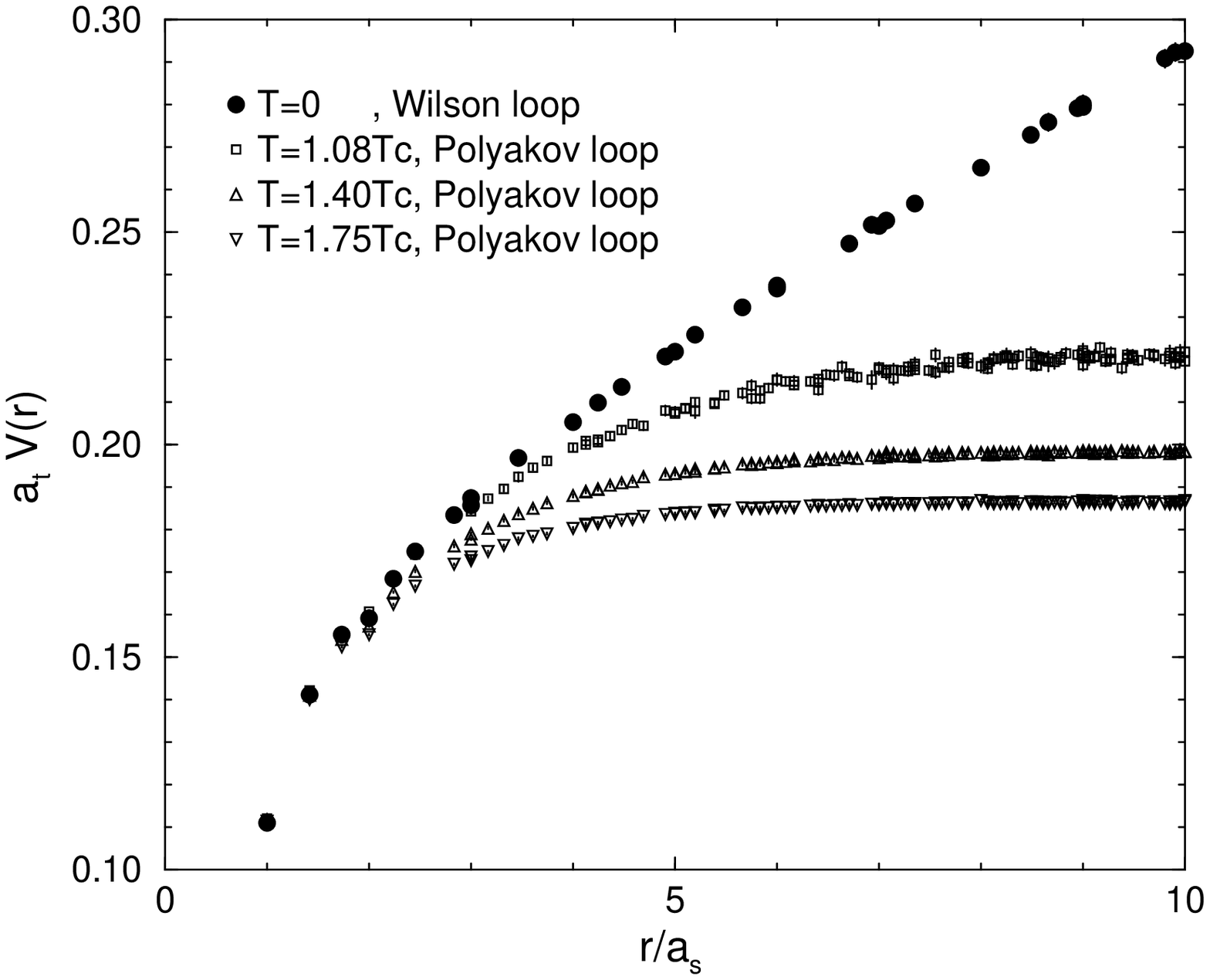}
\vspace{-11mm}
\caption{Color singlet static quark potentials ,$V_{sing}$ and
 $V_{Wilson}$, below and above $T_c$. $T=0$ result is also shown 
 in both figures.}
\label{fig2}
\vspace{-7mm}
\end{figure}

Figure \ref{fig2} shows the result of $V_{sing}, V_{Wilson}$ below 
and above $T_c$ respectively.
Above $T_c$, unfortunately, we fail to extract the $V_{Wilson}$ and 
present only $V_{sing}$ in the bottom panel.
We note that our results of $V_{ave}$ are consistent with 
previous studies, and draws the same conclusions as
Refs.~\cite{Has86,Mat86}.

We solve the Schr\"{o}dinger equation using the quark potential
parametrized to the form, $V(r)=A_1+A_2/r+A_3\cdot r + A_4\cdot r^2$.
Table~\ref{tab1} summarizes the result for the differences of
binding energies from $T=0$.
Below $T_c$, the CCF analysis found the mass difference
of $-(0-10)$ MeV.
Corresponding results of the potential model
with $V_{sing}$ deviate from the CCF results,
while those of $V_{Wilson}$ agree.
Above $T_c$, $V_{sing}$ can hold a bound state
up to $1.08T_c$, while at higher temperatures cannot,
or at least the wave function become too broad
compared to the spatial lattice size.

\begin{table}[tb]
\caption{Result of the Potential model analysis. The numbers mean 
 difference from $T=0$ result.}
\label{tab1}
\begin{tabular}{lll}
\hline
$T/T_c (N_t)$  & $V_{sing}$ (MeV)& $V_{Wilson}$ (MeV) \\
\hline
$0.88 (N_t=32)$  & $+(71-87)$ & $-(0-2)$\\
$0.97 (N_t=29)$  & $+(50-58)$ & $-(21-23)$\\
$1.08 (N_t=26)$  & $-(170-180)$ & none \\
$1.40 (N_t=20)$  &  unbound & none\\
\hline
\end{tabular}
\end{table}

In summary, the disagreement between the results of spectral function
analysis and the potential model is qualitatively absent
if one uses the color singlet free energy as the static quark potential.
This is the same conclusion as the Bielefeld group \cite{Kar04}.
Detailed analysis of the quark potential indicated that
the potential from the Wilson loop most well reproduces the 
result of spectral function analysis below $T_c$.

\end{document}